\documentclass[prl,twocolumn,10pt,aps]{revtex4-2}
\usepackage[utf8]{inputenc}
\usepackage{textcomp}
\usepackage{amssymb}
\usepackage{natbib}
\usepackage{amsmath}
\usepackage{amsfonts}
\usepackage{graphicx}
\usepackage{subfigure}
\usepackage{mathrsfs}
\usepackage{dcolumn} 
\usepackage[T1]{fontenc}
\usepackage{bm}
\usepackage{color}
\usepackage{xcolor}

\usepackage{cancel}
\usepackage{mathbbol}
\usepackage{epsfig}
\usepackage{units}
\usepackage{esint}
\usepackage{soul} 
\usepackage{url}
\usepackage{afterpage}
\usepackage{hyperref}
\usepackage{braket}
\newcommand{\me}[1]{\left\langle #1 \right\rangle }

\newcommand{\gui}[1]{\textcolor{black}{#1}}
\newcommand{\nic}[1]{\textcolor{black}{#1}}

\begin{document}
	\title{ Fractal nature of high-order time crystal phases} 
	%
	\author{Guido Giachetti$^{1,2}$}
	\author{Andrea Solfanelli$^{1,2}$}
	\author{Lorenzo Correale$^{1,2}$}
	\author{Nicol\`o Defenu$^{3}$}
	\affiliation{$^{1}$SISSA, via Bonomea 265, I-34136 Trieste, Italy}
	\affiliation{$^{2}$INFN, Sezione di Trieste, I-34151 Trieste, Italy}
	\affiliation{$^{3}$Institut f\"ur Theoretische Physik, ETH Z\"urich, Wolfgang-Pauli-Str. 27 Z\"urich, Switzerland}
	\begin{abstract}
	\noindent
	
Discrete Floquet time crystals (DFTC) are characterized by the spontaneous breaking of the discrete time-translational invariance characteristic of Floquet driven systems. In analogy with equilibrium critical points, also time-crystalline phases display critical behaviour of different order, i.e., oscillations whose period is a multiple $p > 2$ of the Floquet driving period. Here, we introduce a new, \gui{experimentally-accessible,} order parameter which is able to unambiguously detect crystalline phases regardless of the value of $p$ and, at the same time, is a useful tool for chaos diagnostic. This new paradigm allows us to investigate the phase diagram of the long-range (LR) kicked Ising model to an unprecedented depth, unveiling a rich landscape characterized by self-similar fractal boundaries. Our theoretical picture describes the emergence of DFTCs phase both as a function of the strength and period of the Floquet drive, capturing the emergent $\mathbb{Z}_p$ symmetry in the Floquet-Bloch waves. 
	\end{abstract}
	\maketitle 
\noindent
\emph{Introduction:} The efficacy of technological application of quantum mechanics, such as reliable quantum communication\,\cite{GisinRevModPhys2002}, high-precision quantum metrology\,\cite{GiovannettiNaturePhoton2011}, and fault-tolerant quantum computation\,\cite{PreskillQuantum2018}, depends on the capability of preserving systems out-of-equilibrium, evading the detrimental effects of thermalization, which naturally leads to the loss of locally stored quantum information. Accordingly, much theoretical and experimental effort has been devoted to the study of out-of-equilibrium phenomena\,\cite{PolkovnikovRevModPhys2011,Zhang2017, JurcevicPRL2017,grifoni1998} including, among the others, 
thermalization of isolated quantum many-body systems\,\cite{Rigol2008,KaufmanScience2016,Mori2018,d2016quantum}, dynamical phase transitions\,\cite{HeylPRL2013,Heyl2018, Zunkovic18, markus2017universal, Halimeh2017,correale2021changing} and, finally, the celebrated Discrete Floquet Time Crystals (DFTC)\,\cite{WilczekPRL2012,sacha2015modeling,ElsePRL2016, ElseAnnRevCondMatPhys2020, bordia2017periodically,zhang2017observation,choi2017observation,rovny2018observation}.

The latter are systems where the discrete time-translation symmetry, encoded in the periodically driven Hamiltonian $H(t) = H(t+T)$, is spontaneously broken. 
More precisely, in such a driven system a DFTC phase exists if, taken a class of states $\ket{\Psi}$ with short-ranged connected correlations\,\cite{ElsePRL2016}, it always exists an observable $\hat{O}$ such that the time-evolved expectation value in the thermodynamic limit $N\to\infty$,
\begin{align}
	O(t) = \lim_{N\to\infty}\langle\Psi(t)|O|\Psi(t)\rangle,
\end{align}
satisfies the following conditions\,\cite{RussomannoPRB2017}:
\begin{enumerate}
	\item	Time-translation symmetry breaking: $O(t+T)\neq O(t)$, although  $H(t) = H(t+T)$. This is equivalent to have long-range (LR) correlated Floquet eigenstates of the propagator $U_F = U(t+T,t)$\,\cite{ElsePRL2016}.
	\item Rigidity: $O(t)$ must display periodic oscillations, with some period $\tau$, in a finite and connected region of the Hamiltonian parameters space.
	\item Persistence: in the large system size limit $N\to\infty$, the oscillations of $O(t)$ must persist for infinitely long time.
\end{enumerate}

Conditions 1-3 can not be satisfied by a generic many-body quantum system, in which the presence of an external driving would lead to the relaxation on an infinite-temperature state, ruling out long-lived oscillations. Protecting ordering against relaxation necessitate a mechanism to keep the impact of dynamically generated excitations under control.  \nic{ A natural candidate for the stabilization of pre-thermal phases is strong disorder, which limits the diffusion of  excitations and shall lead to many-body localization (MBL)\,\cite{ElsePRL2016, KhemaniPRL2016, SuracePRB2019,bordia2017periodically,Zhang2017,choi2017observation,roeck2017stability}. Recently, the stability of the MBL phase in the thermodynamic limit has been questioned by state-of-the-art numerical simulations \,\cite{suntajs2020quantum,suntajs2020ergodicity, sels2021dynamical,sels2021quantum,sels2022bath}, generating renewed interest in its phenomenology\,\cite{vidmar2021phenomenology,abanin2021distinguishing,luitz2020absence,crowley2022constructive}. These results make the study of the different mechanisms for pre-thermalization even more pressing, since the traditional arguments on MBL may not apply at large sizes. One possible route to achieve pre-thermal stability in absence of disorder stems from topological protection, which is known to produce long relaxation time even in presence of strong interactions\,\cite{yates2019almost,strong2020lifetime,yates2021strong}. Yet, stable pre-thermal phases whose lifetime grows as the system approaches the thermodynamic limit are only found in presence of non-additive long-range interactions\,\cite{Defenu2021}.}

Then, the possibility of generating a DFTC in clean systems has been studied in the context of LR interacting models, i.e. models in which the interaction between different lattice sites $\mathbf{i}$, $\mathbf{j}$ decay as a power law $J_{\mathbf{i},\mathbf{j}} \sim |\mathbf{i}-\mathbf{j}|^{-\alpha}$. LR systems have sparked a lot of attention recently, due to the possibility of experimental realizations in atomic, molecular
and optical (AMO) systems\,\cite{haffner2008quantum,lahaye2009physics,saffman2010quantum, ritsch2013cold,bernien2017quantum,monroe2021programmable,mivehvar2021cavity,Pagano2018} and to the fact that they exhibit plenty of unique features both for the classical\,\cite{campa2014physics} and the quantum\,\cite{defenu2021longrange} regime. Indeed, LR interactions are known to alter the universal behaviour of critical systems  at equilibrium\,\cite{defenu2017criticality,giachetti2021berezinskii}, and to generate unprecedented out-of-equilibrium phenomena,  with no short-range counterpart, such as novel dynamical phase transitions\,\cite{defenu2019dynamical,halimeh2020quasiparticle}, defect formation\,\cite{acevedo2014new,hwang2015quantum,defenu2018dynamical,defenu2019dynamical}, anomalous thermalization\,\cite{regemortel2016information}, information spreading\,\cite{tran2020hierarchy,chen2019finite,kuwahara2020strictly} and metastable phases\,\cite{Defenu2021,giachetti2023entanglement}. 

In the context of Floquet driven systems, LR interactions are known to enhance the robustness of collective oscillations\,\cite{lerose2019} and the presence of DFTC phases have been actually established in $\alpha < d$ (strong LR) regime\,\cite{RussomannoPRB2017,SuracePRB2019,Munozarias2022,kelly2021stroboscopic}, while for $\alpha > d$ the oscillations are not persistent in the $N \rightarrow \infty$ limit\,\cite{choi2017observation, rovny2018observation,MachadoPRX2020,ColluraArXiv2021}. 
For a long time, the presence of a DFTC phase of order $p$ (that is, with period $pT$, for integer values of $p$), was thought to be connected with the underlying $\mathbb{Z}_p$ symmetry of the model\,\cite{RussomannoPRB2017,SuracePRB2019}, as the Floquet driving can be engineered in such way that each spin approximately oscillates between the $\mathbb{Z}_p$-connected states.
Nevertheless, high-order DFTCs were recently observed also in systems with only $\mathbb{Z}_2$ symmetry, where the order parameter oscillations display a period $pT$ (with $p>2$) given a fixed driving period $T$\,\cite{PizziNatComm2021,giergiel2018time,kelly2021stroboscopic}.
\nic{Actually, high order DFTC phases remain an elusive feature in the landscape of out-of-equilibrium phenomena, first because they are related to an emergent (rahter than fundamental) symmetry\,\cite{Munozarias2022} and, more importantly, because they lack a generic observable, such as an order parameter, which characterizes their appearance}.

In this Letter we solve \nic{this latter issue by} \gui{proposing a new, experimentally-accessible, order parameter which, allows us to achieve a fully fledged characterization of the DFTC phases (regardless of their order) and of the onset of chaos, only relying on geometric features of the dynamics. In the corresponding dynamical phase diagram} the different high-order DFTC phases are found to exhibit a rich pattern of self-similar and fractal structures. In spite of this complexity, we are able capture quantitatively its salient features. We prove that our analysis \gui{is robust against the short-range perturbations, which do not alter the main features of the phase diagram, and finite size effects. In particular,} by performing extensive numerical simulations, 
we find that the emergent $\mathbb{Z}_p$ symmetry is present also at the level of the Floquet eigenstates, and can be interpreted as Bloch superposition of $p$-localized semi-classical states.\\

\textcolor{black}{\emph{The order parameter:} To set the stage, let us consider a generic family driven Hamiltonian $H(t)$, such that $H(t)+H(t+T)$, defined as a function of a parameter (or a set of parameters) $\Lambda$. 
Let us denote $O_n(\Lambda) = \braket{\Psi(nT)|O|\Psi(nT)}$ the average at stroboscopic times of the operator $\hat{O}$ that we aim to use to detect the possible time-crystalline behaviour, for a fixed $\Lambda$. Then, we define the following quantity:
\begin{equation} \label{zeta}
    \zeta^2  = \frac{1}{n_{\rm max}} \sum^{n_{\rm max}}_{n=0} \left[ O_{n} (\Lambda + \delta \Lambda) - O_{n} (\Lambda) \right]^2   \ ,
\end{equation}
in the limit of $\delta \Lambda \rightarrow 0$, $n_{\rm max} \rightarrow \infty$ and $n_{\rm max} \ \delta\Lambda = O(1)$ fixed. Intuitively, $\zeta$ measures the robustness of persistent oscillations of $O_n$, with respect to changes in the driving parameter(s) $\Lambda$.} \\

\textcolor{black}{The main claim of this work is that whenever the stroboscopic dynamics of the observable $O_n$ can be univoquely associated to a classical trajectory, $\zeta$ is a period-blind order parameter, which identifies the DFTC phases independently of their order $p$. This is the case, for example, of fully-connected systems, where the dynamics of any permutationally invariant operator $O$ becomes effectively classical and two-dimensional, within the general theory of Ref.\,\cite{Sciolla2011}. When such system are subject to a periodic force, the corresponding two-dimensional phase space is made-up of a mixture of regular islands, called resonances, and chaotic regions divided by "separatrix" orbits, as foreseen by the Poincar\'e-Birkhoff theorem\,\cite{poincare1912theoreme,birkhoff1913proof}. Within this simple framework, a DFTC phase of order $p$ are known correspond to a periodic hopping of the stroboscopic dynamics of $O_{n}(\Lambda)$ between $p$ resonances, superimposed to a small modulation\,\cite{Munozarias2022,kelly2021stroboscopic,PizziNatComm2021}, so that the distance $|O_{n}(\Lambda+\delta\Lambda)-O_{n}(\Lambda)|$ between trajectories close in parameter space remains finite, saturating to a small value set by the size of resonances. In the opposite chaotic region, $O_{n}(\Lambda)$ and $O_{n}(\Lambda+\delta\Lambda)$ spread uniformly outside the resonances and become uncorrelated on a time-scale $\ln |\delta \Lambda|^{-1} \sim \ln n_{\rm max} \ll n_{\rm max}$, so that the order parameter saturates $\zeta \sim \braket{O(\Lambda)}_{cl}^2$, typically larger than in the DFTC phase, where the brackets $\braket{\cdot}_{cl}$ stand for a uniform classical average over the phase space. In between we observe a third possible behaviour: a periodic dynamics $O_n(\Lambda)$ with a non-rational and strongly $\psi$-dependent period. We refer to the latter as quasi-periodic phase and is typically associated with KAM tori\,\cite{kolmogorov1979conservative,arnold2009proof,moser1962invariant}: as the dynamics is integrable, $|O_{n}(\Lambda+\delta\Lambda)-O_{n}(\Lambda)|$ grows linearly in $n$, and $\zeta$ converges to values whose magnitude is between the values retrieved in the DFTC and chaotic phases, respectively.} \textcolor{black}{As the DFTC and the quasi-periodic phase corresponds to classical trajectories which are topologically distinct, in general $\zeta$ will} \textcolor{black}{exhibit a finite jump between the DFTC and the quasi-periodic phase, in turn distinguishing the two phases apart precisely.\\}

\textcolor{black}{The interplay between the resulting three phases is not expected to be an exclusive feature of mean-field models: all of our discussion can be extend to the inclusion of a weak short-range perturbation or long-range models, where the dynamics can still be rationalized as a single classical trajectory embedded in a self-generated bath of dynamical spin-waves, following the general theory of Refs.\,\cite{Lerose2018chaotic,Lerose2019impact}.}
\emph{The model:} Let us consider the case of a chain of $N$ spin $1/2$ particles, interacting through the Floquet-driven LR Ising Hamiltonian: 
\gui{\begin{equation}
\label{lr_ham}
    H = -\frac{J}{2 N_{\alpha}} \sum_{i > j} \frac{\hat{\sigma}^i_x \hat{\sigma}^{j}_x}{|i-j|^\alpha} + h(t) \sum_i \hat{\sigma}^i_z -\lambda\sum_{i} \hat{\sigma}^x_i\hat{\sigma}^x_{i+1}
\end{equation}}
where $\alpha < 1$,  $\hat{\sigma}^{i}_x$,$\hat{\sigma}^{i}_y$,$\hat{\sigma}^{i}_z$ are the Pauli operators relative to the lattice site $i$, $h(t)$ is a driving periodic with period $T$ and $N_{\alpha} = \sum_{j \neq 0} |j|^{- \alpha}$ is the Kac scaling factor needed in order to have an extensive energy\,\cite{campa2014physics}. At $t=0$, the system is initialized in the ground state of the $h(t)=0$ Hamiltonian, $\textcolor{black}{\ket{\Xi_0}} = \ket{\rightarrow \dots \rightarrow}$, \gui{with $\hat{\sigma}_x \ket{\rightarrow} = \ket{\rightarrow}$}.  Without loss of generality we fix the energy scale such that \gui{$\lambda+J=1$}.

%
In this letter, we take into account the kicked dynamics 
\begin{equation}
    h(t) = \psi \sum^{\infty}_{n=1} \delta(t-nT) \ ,
\end{equation}
($\psi$ being the parameter which determine the strength of the driving). \textcolor{black}{Said $m_a (t) = \frac{1}{N} \sum_{j} \me{\hat{\sigma}^j_a}$ (where $a = x,y,z$), the components of the magnetization of the system we can choose $m_x$ and $\psi$ to play the role of the observable $O$ and the control parameter $\Lambda$ respectively in Eq.\,\eqref{zeta}.} \\

\begin{figure*}
    \centering
    \includegraphics[width=\linewidth]{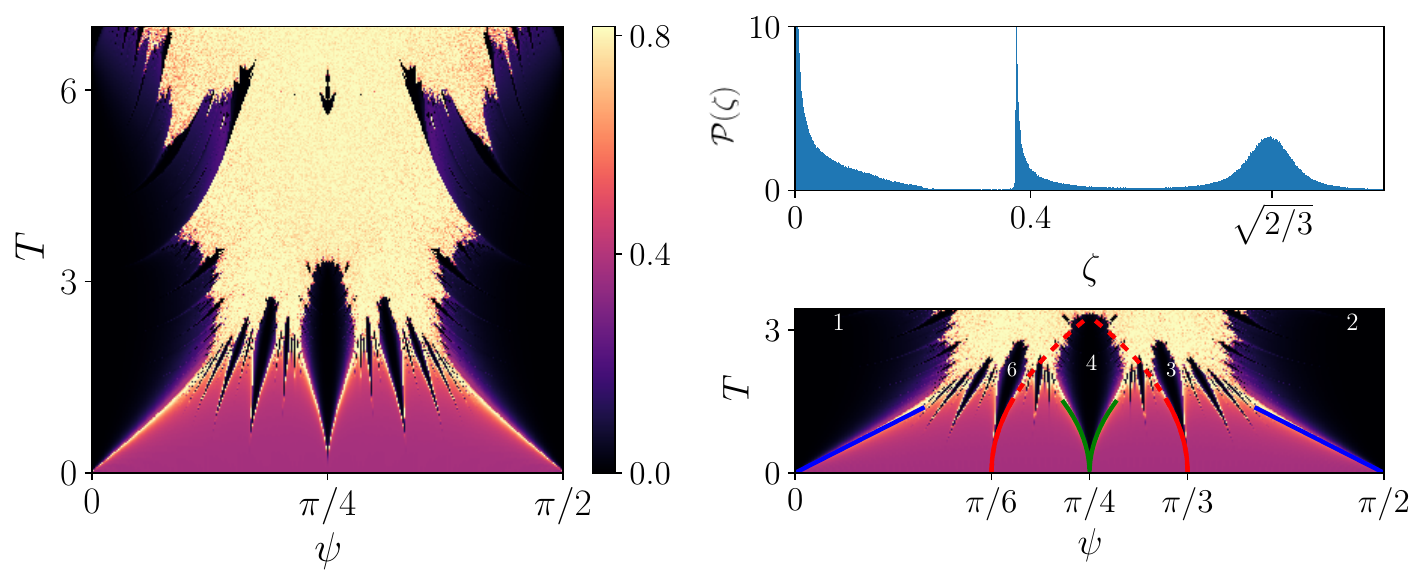}
    \caption{\textit{Left panel:} Color plot of the order parameter $\zeta$ as a function of the amplitude $\psi$ and the period $T$ of the driving, saturated at the value $\zeta = \sqrt{2/3}$, with $n_{\rm max} = 300$, $\delta \psi = 1.6 \cdot 10^{-3}$. \textit{Top-right panel:} \gui{Histogram of the} occurrence \gui{$\mathcal{P}(\zeta)$} of $\zeta$ \gui{within the parameter region of the left panel, normalized at one}. The gap between the DFTC phase, \textcolor{black}{$\zeta \lesssim 0.25$ (and peaked around $\zeta = 0$), and the quasi-periodic one, $\zeta \gtrsim 0.36$,} is apparent. On the right, the profile of the Gaussian distribution around $\zeta = \sqrt{2/3}$, characteristic of the chaotic phase. \textit{Bottom-right panel:} Detail of the $T<3$ region of the phase diagram. The order of the principal DFTC phases is indicated, along with the theoretical prediction for small $T$ of the boundaries between the phases blue, red, green solid lines respectively for the $p=1,2$, $p=4$, $p=3,6$ islands (whose exact expression is given in the Supp. Mat.\,\cite{SM}). The prolongation of the boundaries of the $p=3$, $p=6$ islands (dashed red line) gives a good estimate of the onset of chaos, which disrupts the time-crystal phases at large $T$.}
    \label{Fig1}
\end{figure*}

\emph{Mean-field results:} \gui{First, we consider the fully-connected $\alpha=\lambda=0$ case in thermodynamic limit. \textcolor{black}{Said $\mathbf{m} \equiv (m_x,m_y,m_z)$ and $m_{a,n} = m_a(nT)$} with $a = x,y,z$, the dynamics of $\mathbf{m}_n \equiv \mathbf{m} (nT)$ is determined by the mean-field map\,\cite{sciolla2013quantum}:}
\begin{equation} \label{map}
    \mathbf{m}_{n+1} = f(\mathbf{m}_n)  \equiv R_z (2 \psi) R_x (-m_{x,n}T) \mathbf{m}_{n} \ ,
\end{equation}
with $\mathbf{m}_0 = (1,0,0)$ and where $R_{x,y,z} (\xi)$ is the rotation matrix of an angle $\xi$ around the corresponding axis (see Supp. Mat.\,\cite{SM} and Ref.\,\cite{Munozarias2022}). 
\textcolor{black}{The general picture valid for fully-connected models can be made explicit here, as shown in Supp. Matt. \cite{SM}; in particular, due to the constraint $\mathbf{m}^2 =1$ in the chaotic phase $\zeta^2$ is peaked around $\zeta^2 = 2/3$.  The predictions are numerically checked in Fig.\,\ref{Fig1} (top right-panel).} 

The value of $\zeta$ as a function of $\psi$ and $T$ is shown in Fig.\,\ref{Fig1} (left-panel). The resulting structure is strikingly complex and convoluted. The phase diagram is symmetric around the $\psi = \pi/4$ axis as a consequence of the dynamical $\mathbb{Z}_2$ symmetry, \gui{a feature that could not have been observed with a $p$-dependent order parameter}. For small $T$ the quasi-periodic phase is prevalent, while small islands of the periodic phase appear around some particular values of $\psi$ which correspond to rational multiples of $\pi$. Initially, the size of these islands grows with $T$ and, as they get closer to each other, chaos start to onset around their boundaries. Finally, all the islands corresponding to a DFTC of order $p>2$ are swallowed by the chaotic phase, the last one corresponding to $p=4$. In correspondence of particular values of the driving period we have a revival of the higher-order DFTC phases, especially visible in correspondence of $p=4$. 

The boundary between the chaotic and the DFTC phase is not smooth: rather, it presents plenty of self-similar patterns which are repeated at smaller and smaller scales. Taking as an example the boundary of the $p=3$ island, the numerical estimate of its Minkowski - Bouligand dimension\,\cite{Strogatz2018}, $d_{MB}$, gives the value
\begin{equation} \label{dmb}
    d_{MB} \approx 1.4(2) > 1
\end{equation}
(see Supp. Mat.\,\cite{SM}) indicating a fractal nature. This is compatible with the estimate of $d_{MB}$ in the entire phase diagram. The appearance of fractal scaling for time crystalline phase boundaries draws a direct analogy with similar phenomena in traditional critical systems, especially percolation, self-avoiding random walks and Potts model\,\cite{hastings2002exact,duplantier2000conformally}, where a rigorous connection between conformal invariance and stochastic evolution has been verified\,\cite{kager2004guide,cardy2005sle}.  
As already noticed in Ref.\,\cite{Munozarias2022}, the formation DFTC islands can be understood within the formalism of area-preserving maps \,\cite{mackay1982renormalisation}, and in particular can be linked to the existence of Arnold tongues \,\cite{vulpiani2009chaos,collado2021emergent}, \nic{which also appear in the pre-thermal time-crystal phase of driven $O(\mathcal{N})$-symmetric models\,\cite{natsheh2021critical,natsheh2021critical2}.}

\textcolor{black}{As shown in Supp. Matt. \cite{SM}, the particular form of Eq.\,\ref{map}, allows us to probe our general picture, and even to reproduce the main feature of the phase-diagaram for small $T$. Indeed, as for $T=0$ the  map \eqref{map} is nothing but a rotation of angle $\psi$ around $z$, any $\psi = \psi_r \equiv r \pi$, with $r=q/p$ (and $q$ coprime with $p$) corresponds to a fixed point of the iterated map $f^{\Tilde{p}}$ with $\Tilde{p} = p$,$\Tilde{p} = p/2$ for odd and even $p$ respectively (due to the $\mathbb{Z}_2$ symmetry of the model). For $\psi - \psi_r \equiv \delta \psi \ll \pi/p$ and $T \ll 1$ the dynamics of $f^{\Tilde{p}}$ is slowed down and it can be described as an Hamiltonian flow generated by 
\begin{equation} \label{hamphiI}
        H(\phi, I) = 2 \frac{\delta \psi}{T} I  - \frac{1}{4} (1 - I^2) \left( 1 + a_r \cos (2 \phi - 2 \psi_r) \right) \ ,
\end{equation}
where $\phi$ is the azimuthal angle of $\mathbf{m}$ and $I = m_z$ its conjugate momentum. Let us notice that $r$ enters in Eq.\,\eqref{hamphiI} only through the coefficient $a_r$, (whose exact expression is given in the Supp. Matt.\,\cite{SM}) thus accounting for the presence of self-similar structures. Studying the different topology of the trajectories of Eq.\,\eqref{hamphiI} we are able to estimate the boundaries DFTC islands (their expressions are in the Supp. Matt.\,\cite{SM}), which are in agreement with the numerics, see Fig.\,\ref{Fig1} (bottom-right panel).} 

According to the  Chirikov criterion \,\cite{chirikov1979universal}, the value of $T$ at which two of these curves intersect can be taken as an estimate of the threshold $T_{*}$ beyond which the chaos takes over: this gives $T_{*} = (12 \pi^2)^{1/4} \approx 3.299$ for onset of chaos in the $p=4$ island, which is in excellent agreement with the numerics. 
%
%
\\\\
\gui{\emph{Beyond the mean-field:} The analysis
of the DFTC phases we reported in Fig.\,\ref{Fig1} can be straightforwardly extended both beyond the fully-connected case and to account for finite sizes.
To test the robustness of our results for finite $N$, we checked numerically the structure of the higher-order Floquet eigenstates $\ket{\eta_m }$, in the fully-connected case. The results of our analysis are shown in Fig.\,\ref{Fig2} (a): by introducing a coherent state representation\,\cite{auerbach2012interacting} the Floquet eigenstates in the $p=4$ DFTC phase appear clearly localized around four $\mathbb{Z}_4$ symmetric points, while this is no longer the case in the quasi-periodic phase. As explained in the Supp. Mat.\,\cite{SM} (see also \cite{RibeiroPRE2008} for the details on the numerics), this behavior can be explained semi-classically: close to a resonance, the Floquet evolution can be interpreted as a hopping between $p$ adjacent wells in the classical phase space\,\cite{giergiel2018time}, so that the Floquet eigenstates have the form of tight-binding Bloch wavefunctions. A similar behavior for the $p=2$ case (around $\psi = \pi/2$) has already been observed in Ref.\,\cite{RussomannoPRB2017}.
We also study the stability of the phase diagram of the inclusion of finite-range perturbation, implemented by posing by finite value of either $\alpha$ or $\lambda$, within the framework of non-equilibrium spin-wave theory\,\cite{Lerose2018chaotic,Lerose2019impact,SM}: the plots in Fig.\,\ref{Fig2} (b) and (c) show that the structure of the low-T DFTC regions is not qualitatively altered by the perturbation, while the DFTC island around $\psi=\pi/4$ and $T=6$ disappears for sufficiently strong values of $\alpha$ or $\lambda$.
These observations the validity of our description, beyond the mean-field limit.}
\\
\begin{figure*}[t!]
    \centering
    \includegraphics[width=1.05\linewidth]{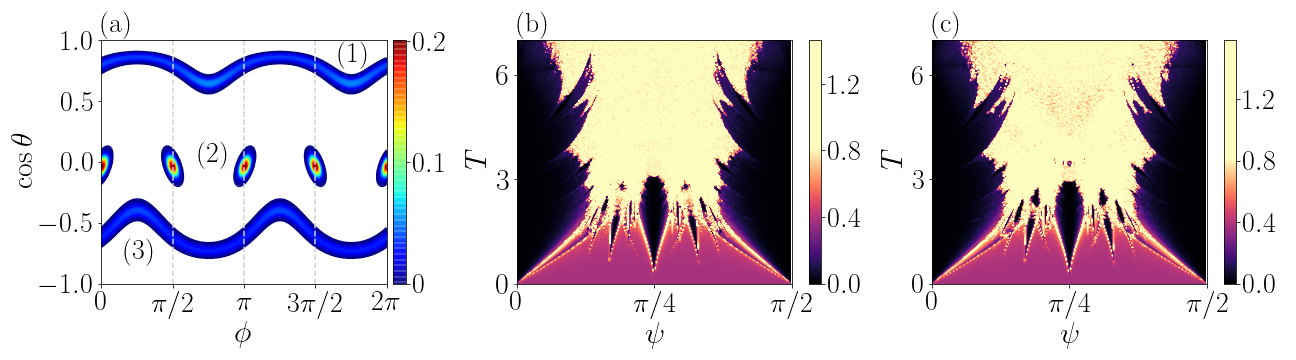}
    \caption{(a): Color plot of the overlap $| \braket{\Omega_{\theta, \phi} | \eta_m}|^2$ between the spin coherent state $\ket{\Omega_{\theta,\phi}}$ and different Floquet eigenstates $\ket{\eta_m}$, for $N=800$, $\psi = \pi/2 + 0.01$, $T=1$. The disconneccted structure of the eigenstate (2), corresponding to the $p=4$ DFTC,  is apparent. 
    (b): Color plot of the order parameter $\zeta$ as a function of the amplitude $\psi$ and the period $T$ of the driving, obtained for $\alpha=0.3$, $\lambda=0$ and the same simulation parameters of Fig.\,\ref{Fig1} (left panel).
    (c): Same color plot of panel (b), obtained for $\alpha=0$, $\lambda=0.06$.
    }
    \label{Fig2}
\end{figure*}
 

\emph{Conclusions:} In this Letter 
we introduced a new order parameter $\zeta$ able to unambiguously detect higher-order Discrete Floquet Time-Crystal (DFTC) in clean long-range (LR) systems. The pressing need for this study was generated by the recent depiction of high order DFTC phases\,\cite{Munozarias2022,PizziNatComm2021}. 
\nic{We expect $\zeta$ to become a universal tool for detecting DFTC, beyond the current model},
\textcolor{black}{as it exploits the connection between DFTC and Poincar\'e-Birkhoff theorem\,\cite{poincare1912theoreme,birkhoff1913proof} (see\,\cite{Wisniacki2011Poincare} for the quantum counterpart) within the mean-field regime (see Refs.\,\cite{Munozarias2022,kelly2021stroboscopic} for a detailed analysis of the $p$-spin model case). }
\textcolor{black}{
Choosing as a paradigmatic example the kicked LMG model, we are able to draw a new phase diagram, featuring self-similar structures with non-integer, fractal, dimension and to explain quantitatively this phenomenon in terms of a an effective Hamiltonian map with renormalized couplings. While our picture become exact at $\alpha = 0$, $N = \infty$, we verified its roboustness for finite size and finite $\alpha$, and in presence of competing short-range interactions.}




We stress the fact that our newly introduced observable $\zeta$ is easily accessible in the experiments, so that our predictions on the phase diagram may be tested
e.g. in NMR experiments on driven, ordered systems\,\cite{rovny2018observation}. Our characterization of the higher-order, stable DFTC phases could thus pave the way for a series of advancement in the field of quantum technologies. 

\emph{Acknowledgements:} This research was funded in part by the Swiss National Science Foundation (SNSF) [200021\_207537]. This work is supported by the Deutsche Forschungsgemeinschaft (DFG, German Research Foundation) under Germany’s Excellence Strategy EXC2181/1-390900948 (the Heidelberg STRUCTURES Excellence Cluster).


%

%

\begin{widetext}
\begin{appendix}

\title{Supplementary Material: Fractal nature of high-order time crystal phases} 

\maketitle

\section{Derivation of Eq. (4)}
\noindent
Here we revise the derivation of the dynamic map of Eq. (4) of the main text, i.e. the evolution equation of the magnetization $m_a(t) = \sum_j \braket{\hat{\sigma}_a^j } /N$, at stroboscopic times  $t_n = nT$, in the thermodynamic limit, $N \rightarrow \infty$.

First we notice how, due to the impulsive nature of the magnetic field $h(t)$, the Floquet propagator can be written as the product of two different operators:
\begin{equation} \label{Uf}
    U_F = e^{- 2 i \psi \hat{S}_z} e^{i J T \hat{S}_x^2/N} \ , 
\end{equation}
where we introduced the global spin operators
\begin{equation}
    \hat{S}_a = \frac{1}{2} \sum_{j} \hat{\sigma}_a^j \ ,
\end{equation}
with $a = x,y,z$. Being $\hat{S}_x$ the generator of $O(3)$ rotations around the $x$ axis, the kick term $\exp(-2i\psi \hat{S}_z)$ in Eq.\eqref{Uf} acts on the observable $\mathbf{m}$ simply as a rotation around the $z$-axis. On the other hand, the other term describes the evolution over one period $T$ of $\mathbf{m}$ induced by the second term on the r.h.s of Eq.\,\eqref{Uf}. The Heisenberg equations of motion corresponding to such evolution for the operators $\hat{S}_a$ are: 
\begin{equation} \label{HeisS}
    \left\{
    \begin{split}
        \frac{d}{dt}\hat{S}_x &= 0 \ ,  \\
        \frac{d}{dt}\hat{S}_y &=  \frac{J}{N} \left(\hat{S}_x \hat{S}_z + \hat{S}_z \hat{S}_x \right) \ ,  \\
        \frac{d}{dt}\hat{S}_z &=  -\frac{J}{N} \left(\hat{S}_x \hat{S}_y + \hat{S}_y \hat{S}_x \right) \ .
    \end{split}
    \right.
\end{equation}
According to the general theory developed in Ref. \cite{sciolla2013quantum}, for $N\to\infty$ the spin-spin correlation become negligible, that is $\braket{\hat{S}_a\hat{S}_b}\simeq \braket{\hat{S}_a}\braket{\hat{S}_b}$, so that Eqs.\,\eqref{HeisS} become a closed set of equations for $m_a = 2 \me{S_a}/N$, namely: 
\begin{equation} \label{HamM}
    \left\{
    \begin{split}
        \dot{m}_x &= 0 \ ,  \\
        \dot{m}_y &=  J m_x m_z \ , \\
        \dot{m}_z &=  - J m_x m_y \ .
    \end{split}
    \right.
\end{equation}
In turn this results after a time $T$ in a (clockwise) rotation around the $x$-axis, of angle $J T m_x(t)$. The $\mathbb{Z}_2$ symmetry of the model is instead encoded into the dynamical symmetry $\psi \rightarrow \psi + \pi/2$, $\mathbf{m}_n \rightarrow R_z (\pi n) \mathbf{m}_n$ of Eq.\,\eqref{HamM}.
Posing $J=1$, the overall effect of the Eq. \eqref{Uf} on our observable $\mathbf{m}$ is the one of Eq.(4) of the main text.

\section{Hamiltonian chaos and the order parameter $\zeta$}
In this Section we will derive in detail some properties of the order parameter $\zeta$, in the context of the theory of symplectic maps. 

Indeed, let us notice that the map of Eq. (4) inherits a Hamiltonian structure, which is manifest in the fact that the area of any region on the sphere $\mathbf{m}^2 = 1$ is preserved by the action of $f(\mathbf{m})$. In terms of the usual polar coordinates along the $z$ axis,  $\mathbf{m} = (\sin \theta \cos \phi, \cos \theta \cos \phi, \cos \theta)$, the area element can be written as $dS = d \cos \theta d \phi$, so that $\phi$ and $I = \cos \theta$ are natural canonical conjugate variables for our system. This is coherent with the picture developed in Ref.\cite{sciolla2013quantum}, and can be intuitively understood by thinking of $I$ as the $z$ component of the angular momentum, and $\phi$ the coordinate corresponding to rotation around the $z$ axis. 

In particular, in the limit $T=0$, the map becomes 
\begin{equation} \label{AA}
\begin{split}
    I_{n+1} &= I_{n} \ , \\
    \phi_{n+1} &= \phi_n + 2 \psi \ ,
\end{split}
\end{equation}
with $I_0 = 0$, $\phi_0 = \pi/2$. This correspond to the stroboscopic section of an integrable dynamics, $\phi$ and $I$ playing the role of an angle-action pair. In terms of the Floquet phases introduced in our paper, this implies a quasi-periodic evolution of $\mathbf{m}_n$, with period $\pi/\psi$. 

As a small $T$ is switched on, it can be treated as a perturbation to the map \eqref{AA}. In this case, the fate of the system is described by the Kolmogorov-Arnold-Moser theorem \cite{kolmogorov1979conservative,arnold2009proof,moser1962invariant}, according to which the torus $I = \rm const$ is only deformed as long as the corresponding frequency is not resonant. In turns this means that the quasi-periodic phase survives as long as $\psi$ is not close to a rational multiple of $\pi$. In the case of $\psi$ close to a $q:p$ resonance, i.e. $\psi = \psi_{r} \equiv r \pi$, with $r=q/p$ and $p$ and $q$ coprime integers, according to the Poincare-Birkhoff theorem, pairs of elliptic and unstable fixed points are expected to arise. In this case the action of the $p$-iterated map $f^p(\mathbf{m})$ individuates different regions in the phase space $(I, \phi)$, corresponding to different possible behaviors of $\mathbf{m}_n$. In particular, if $(I_0, \phi_0)$ is far from the fixed points, we have a rotation dynamics, with $\phi$ growing from $0$ to $2 \pi$, and we have a quasi-periodic behavior. If $(I_0, \phi_0)$ is close to one of the centers instead, we have a libration dynamics, with $\phi$ oscillating around a finite value. As a consequence, $m_{n+p}$ remains close to $m_n$, and we have a DFTC phase. At the boundaries between this two regions a chaotic region is expected to arise which, as $T$ increases, grows and possibly swallows the regular ones. 

Let us analyze the consequence of this picture on the order parameter $\zeta$: to consider different values of the amplitude, $\psi$, $\psi + \delta \psi$ it means that we are considering to nearby initial conditions on the phase space. Both in the DFTC phase and in the quasi-periodic one the evolution is not chaotic, so that the two trajectories will diverge
polynomially in time. As a consequence
\begin{equation} \label{zetareg}
    \zeta^2 = \frac{1}{n_{\rm max}} \sum^{n_{\rm max}}_{n=0} \left( m_{x,n} (\psi + \delta \psi) -  m_{x,n} (\psi) \right)^2 \sim \frac{\ell}{n_{\rm max}} \sum^{n_{\rm max}}_{n=0} \delta \psi^2 n^2   \sim \ell (\delta \psi n_{\max})^2 \ ,
\end{equation}
where $\ell$ depends on the the average distance between two randomly chosen points of the two nearby trajectories. While the r.h.s. of Eq.\,\eqref{zetareg} remains finite and $O(1)$ as $n_{\rm max} \rightarrow \infty$, $\delta \psi \rightarrow 0$,  $\ell$ jumps discontinuously as we pass from the libration regime (corresponding to a DFTC phase) and the rotation one (corresponding to a quasi-periodic phase). In particular, as we approach the the fixed point of the iterate map, i.e. in the regime in which the micro-motion is becomes negligible, $\zeta \rightarrow 0$, so that $\zeta$ is able to quantify how far away is our system from the pure crystalline regime. Let us notice, however, how the other values of $\zeta$ in this two phases are not universal, as they depend on the choice of $ \lim_{\delta \psi \rightarrow 0} n_{\rm max} \ \delta \psi$.


In the chaotic phase, instead, the trajectories diverge exponentially, so that after a time-scale $n_{\rm max} \sim -\log(\delta \psi)$ the memory of the initial condition is lost. In this case we can assume each $m_{x,n} (\psi)$ and $m_{x,n} (\psi + \delta \psi)$ to be drawn from a set of equally distributed random variables with zero mean. As a consequence, according to the central limit theorem, $\zeta^2$ is distributed as a Gaussian around the value
\begin{equation}
    \me{\zeta^2} = 2 \me{m_{x}^2} 
\end{equation}
with a variance $O(n_{\rm max}^{-1})$. Assuming furthermore the distribution of the three components to be isotropic, and taking into account the constraint $\mathbf{m}^2 = 1$, we have 
\begin{equation}
    \me{m_{x}^2} = \frac{1}{3} \me{\mathbf{m^2}} = \frac{1}{3} \ ,
\end{equation}
so that $\me{\zeta^2} = 2/3$. 


\section{Non-integer dimension of the chaotic phase boundary}
\begin{figure}
    \centering
    \includegraphics[width=0.45\linewidth]{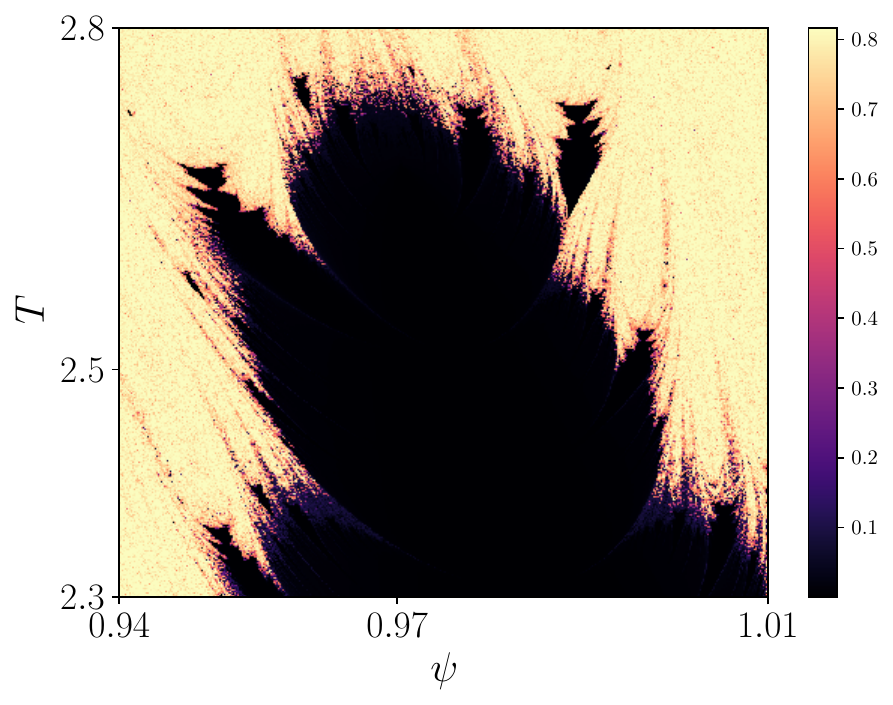} \
    \includegraphics[width=0.45\linewidth]{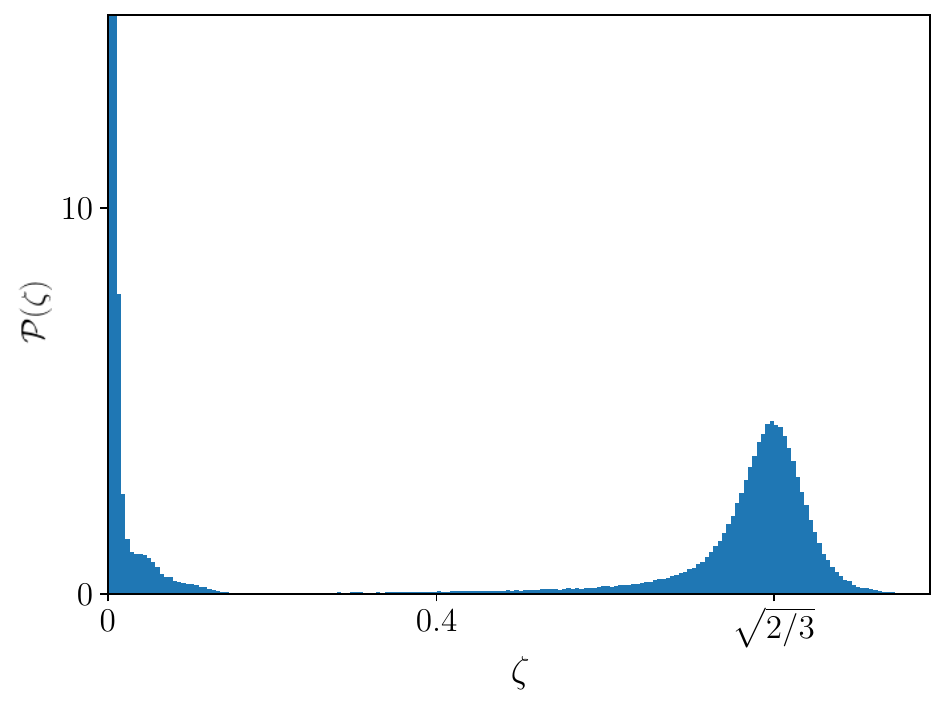}
    \caption{Color plot of the order parameter $\zeta$ saturated at the value $\zeta = \sqrt{2/3}$ with $n_{\rm max} = 300$, $\delta \psi = 10^{-4}$ (left) and corresponding normalized occurrence frequency $\mathcal{P} (\zeta)$ of $\zeta$ for this region of the phase diagram.}
    \label{fig:Fig12}
\end{figure}
The Minkowski-Bouligand, or box-counting, dimension is defined as follows: let us cover our space with and evenly spaced square grid of side $\epsilon$. Said $N(\epsilon)$ the number of boxes which lies on the boundary, then the dimension is defined as
\begin{equation}
    d_{MB} \equiv - \lim_{\epsilon \rightarrow 0} \frac{\ln N(\epsilon)}{\ln \epsilon} \ .
\end{equation}
Since we are interested to the border between the chaotic and the DFTC phases, we restricted ourselves to a region of the phase diagram in which only these phases are present. Here we choose $\psi \in (0.94, 1.01)$, $T \in (2.3, 2.8)$, which corresponds to the edge of the $p=3$ island. The region is shown in Fig.\,\ref{fig:Fig12}, left panel. To precisely define our boundary, we have here to define a threshold $\zeta_{*}$ such that a point with $\zeta > \zeta_{*}$ is considered to belong to the chaotic phase. 

Since for any finite $n_{\rm max}$ the  distribution of $\zeta$ in the chaotic phase around $\zeta  = 0.816$ has a finite width, it is convenient, in this case, to choose $\delta \psi$ such that, $n_{\rm max} \ \delta \psi << 1$. In this regime, indeed, the distribution of $\zeta$ in the DFTC phase is sharply peaked around $0$, and the separation of the two phases more pronounced. If, however, we choose $\delta \psi$ to be too small, the diagnostic of the chaotic region will not be accurate, as $\mathbf{m}_{n} (\psi)$ will not forget its initial condition for $n \sim n_{\rm max}$. We checked that the choice $n_{\rm max} = 300$, $\delta \psi = 10^{-4}$ is close to the optimal one. The separation between the different phases, shown in Fig.\,\ref{fig:Fig12}, right panel, is clear. 

The behavior of $d_{MB}$ as a function of the cutoff $\zeta^{*}$ Fig.\,\ref{fig:Fig34}, left panel: as expected i, $d_{MB}$ shows a very weak dependence on $\zeta^{*}$ within a finite interval, signaling that the two phases are well distinct. The corresponding value of the fractal dimension turns out to be $d_{MB} \approx 1.42$. Repeating the procedure with slightly different values of $n_{\rm max}$ and $\delta \psi$ gives an uncertainty of the order of order $10^{-2}$ on the above result.  

\begin{figure}
    \centering
    \includegraphics[width=0.45\linewidth]{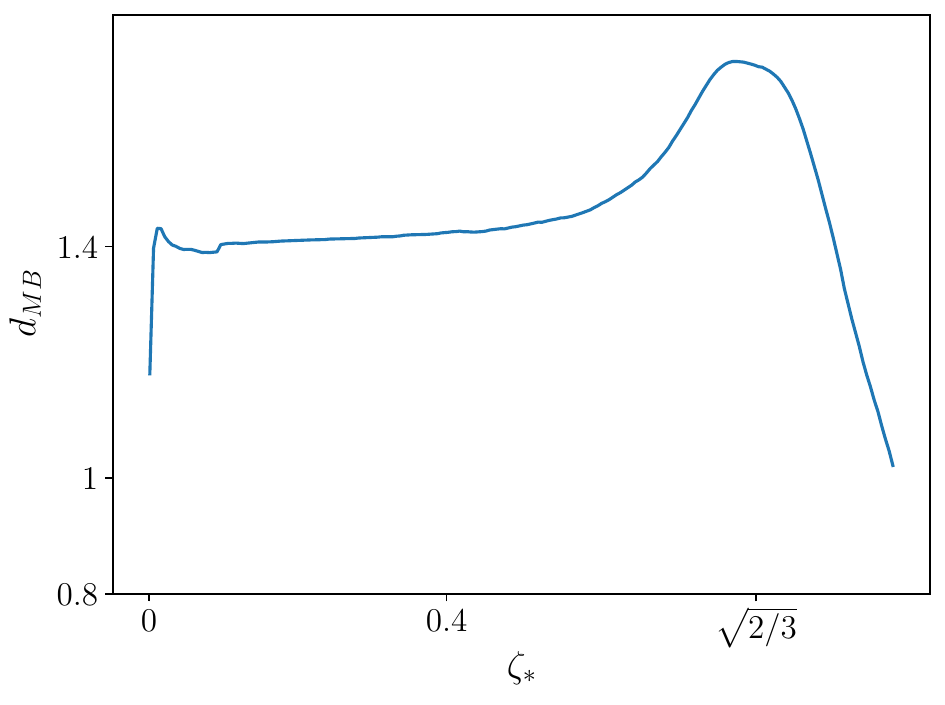} \
    \includegraphics[width=0.45\linewidth]{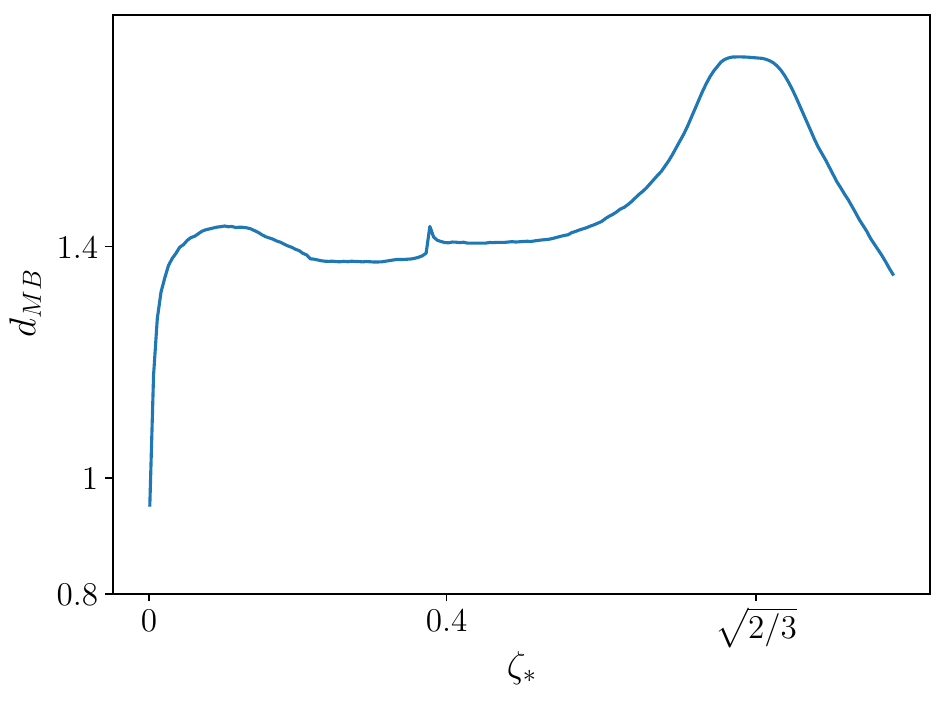}
    \caption{Behavior of the Minkowski-Bouligan dimension $d_{MB}$ of the boundary between the chaotic DFTC phases as a function of the threshold $\zeta_{*}$ chosen, for the case of the restricted area of Fig. (left) and for the case of the whole phase diagram (right).}
    \label{fig:Fig34}
\end{figure}

In Fig. \ref{fig:Fig34} right panel, we reported the behavior of $d_{MB}$ as a function of the threshold $\zeta_{*}$ for the whole phase-diagram of Fig. 1 of the main text. To rule out the effect of the quasi-periodic phase, which is now present, we have to restrict ourselves to the region $\zeta_{*} \gtrsim  0.36$. As a consequence, the resulting estimate is far less accurate, and the dependence of $d_{MB}$ on $\zeta_{*}$ more pronounced. We see, however, that the our previous, local, estimate is fully compatible with this global results.

\section{Small $T$ expansion}\label{APP_analytic estimation}
Let us now derive the results presented in the paper, which are valid for small $T$ and $\psi$ close to $\psi_r = r \pi$, where $r= q/p$, $q$,$p$ being coprime integers. 

First, let us expand the map of Eq. (4) of the main text to the first order in $T$. As in Section B, it is convenient to parametrize the magnetization as $\mathbf{m} = (\sin \theta \cos \phi, \sin \theta \sin \phi, \cos \theta)$ and introduce the canonical coordinates $I = \cos \theta$, $\phi$. We obtain 
\begin{equation} \label{linearized}
    \begin{split}
        I_{n+1} &= I_n - \frac{T}{2} (1 - I^2_{n+1}) \sin 2 \phi_n + O(T^2) \ , \\
        \phi_{n+1} &= \phi_n + 2 \psi + \frac{T}{2} I_{n+1} \left( 1 + \cos 2 \phi_n \right) + O(T^2) \ .
    \end{split}
\end{equation}
Let us notice that the $O(T)$ terms on the r.h.s. of the above equations, $I_{n+1}$ can be replaced by $I_n$ up to higher order corrections. This choice guarantees that our approximation preserves the symplectic structure of the original map. The above is invariant under the discrete translation $\phi_n \rightarrow \phi_n + n \pi$; this is a consequence of the dynamical $\mathbb{Z}_2$ symmetry of the original model. 

Let us now build the equivalent of the map \eqref{linearized} for the generic $\tilde{p}$-iterated map. For $T=0$ we have than that the action of the $\tilde{p}$-iterated map is trivially $\phi_{n+\tilde{p}} = \phi_n + 2 \tilde{p} \psi$, $I_{n+\tilde{p}} = I_n$. Then, up to higher order in $T$ we can write:  
\begin{equation} \label{piterated}
    \begin{split}
        I_{n+\tilde{p}} &= I_n - \frac{\tilde{p} T}{2} (1 - I^2_{n+p}) U_{\tilde{p}} (\psi) \ \sin ( 2 \phi_n + 2 (\tilde{p}-1) \psi) \ , \\
        \phi_{n+\tilde{p}} &= \phi_n + 2 \tilde{p} \psi + \frac{\tilde{p} T}{2} I_{n+  \tilde{p}} \left[ 1 + U_{\tilde{p}} (\psi) \cos ( 2 \phi_n + 2 (\tilde{p}-1) \psi ) \right] \ ,
    \end{split}
\end{equation}
where 
\begin{equation}
    U_{\tilde{p}} (\psi) = \frac{\sin 2 \tilde{p} \psi}{\tilde{p} \sin 2 \psi} \ .
\end{equation}
Being $T$ small, in general the $ 2 \tilde{p} \psi$ term in the r.h.s. of the second equation of Eq\,\eqref{piterated} is going to dominate the evolution so that, we expect to find ourselves in the quasi-periodic phase. However, as $\tilde{p} \psi \sim  k \pi$ for some integer $k$ we have that this terms becomes small as well, signaling a the onset of the Poincaré-Birkhoff mechanism (for $k$ odd, this is due to the $\mathbb{Z}_2$ symmetry, which allows us to reabsorb the term by redefining $\phi_n \rightarrow \phi_n + n \pi$).

Let us put ourselves close to a resonance, i.e. let us consider the limit $\psi = \psi_r + \delta \psi$ with $\delta \psi \ll \pi/p$. If $p$ is odd, the smallest choice of $\tilde{p}$ which makes the term $\tilde{p} \psi$ small is $\tilde{p} = p$; if $p$ is even, however, we have to choose $\tilde{p} = p/2$ (and redefine $\phi_n \rightarrow \phi_n + n\pi$. 
In this limit the Eq.\,\eqref{piterated} becomes 
\begin{equation} \label{firstorder}
    \begin{split}
        I_{n+\tilde{p}} &= I_n - \frac{\tilde{p} T}{2} (1 - I^2_{n+\tilde{p}}) a_r \ \sin ( 2 \phi_n - 2 \psi_r ) \ , \\
        \phi_{n+\tilde{p}} &= \phi_n + 2 \tilde{p} \delta \psi + \frac{\tilde{p} T}{2} I_{n+\tilde{p}} \left( 1 + a_r \cos ( 2 \phi_n - 2 \psi_r ) \right) \ ,
    \end{split}
\end{equation}
where 
\begin{equation}
    a_r = \begin{cases}
    1 \ \ &\text{if} \ p=1,2 \\
    2 (-1)^{\tilde{p}-1} \csc(2 \psi_r) \delta \psi \ \ &\text{if} \ p \geq 3 \ . 
    \end{cases}
\end{equation}
Let us notice how in Eq.\,\eqref{firstorder} the evolution of both $\phi$ and $I$ is now slow, signaling that the $\tilde{p}$-iterated map can be approximated by a continuous flow. In order to do so, we have to redefine the time scale  $\tilde{p} T \rightarrow T$, such that $\phi_{n+ \tilde{p}}, I_{n+ \tilde{p}} \rightarrow \phi_{n+1}, I_{n + 1}$ now by introducing the time step $\Delta t = \tilde{p} T$, and expanding $I_{n+1} = I_n + \dot{I} \ T + O(T^2)$, $\phi_{n+1} = \phi_n + \dot{\phi} \ T + O(T^2)$. We find then
\begin{equation}
    \begin{split}
        \dot{I} &= - \frac{a_r}{2} (1 - I^2) \sin (2 \phi - 2 \psi_r) \ , \\
        \dot{\phi} &= 2 \tilde{p} \frac{\delta \psi}{T} + \frac{1}{2} I \left( 1 + a_r \cos (2 \phi - 2 \psi_r) \right) \ . 
    \end{split}
\end{equation}
In turn, this can has the form of an Hamiltonian flow, generated by
\begin{equation} \label{Ham}
    H(\phi, I) = 2 \frac{\delta \psi}{T} I  - \frac{1}{4} (1 - I^2) \left( 1 + a_r \cos (2 \phi - 2 \psi_r) \right) \ .
\end{equation}
By taking into account our initial condition, namely $\phi(0) = \pi/2$, $I(0) = 0$ we have that our $p$-iterated dynamics takes place along the curve $ H (I, \phi) =  - \frac{1}{4} \left( 1 - a_r \cos (2 \psi_r) \right)$. 

If this curve is bounded between two finite values of the angle $\phi$, the dynamics will circle around a fixed point, signaling that we are within the time crystalline phase. If, instead, the curves cover the all $\phi \in [0, 2 \pi]$ interval, we have a quasi-periodic motion. Finally, close to the separatrix between this two cases, chaos is expected to arise, so that the boundary between these two regimes in terms of $T$, $\phi$ can be taken as an estimate of the edge of the DFTC for small $T$. 

For $p=1,2$, this criterion gives the condition $T= 4 |\delta \psi|$. For $p >2 $, instead, at the lowest order in $\delta \psi$ we find, for any $\psi_r < \pi/2$, the condition  
\begin{equation} \label{Tpsi}
    T^2 = \begin{cases}
     8 \  \tilde{p}^2 \tan \psi_r \ |\delta \psi| \ \ \ &\text{if} \ \delta \psi < 0 \ , \\
     8 \ \tilde{p}^2 \cot \psi_r \ |\delta \psi| \ \ \ &\text{if} \ \ \delta \psi > 0 \ . \\
    \end{cases}
\end{equation}
Let us notice how Eq.\eqref{Tpsi} captures both the non-analytic behavior of the boundary of the DFTC and its lack of symmetry around the resonant value $\psi_r$ for $r \neq 1/4$. However, in order for the result to be predictable at the quantitative level we have to impose that the term $\delta \psi/T$ in Hamiltonian \eqref{Ham} to be small, this implies that the steepest curve between the two of Eq.\,\eqref{Tpsi} is not a good approximation of the boundary in the whole region $|\delta \psi| \sim \pi/p$, and its coefficient is not reliable. 

For $r = 1/4$ ($\tilde{p} =2$) Eq.\,\eqref{Tpsi} gives
\begin{equation}
    T^2 = 32 | \delta \psi| \ ,
\end{equation}
while for $r =1/6$ and $r =1/3$ ($\tilde{p} = 3$) we find that the steepest edge grows respectively as
\begin{equation}
    T^2 = \pm 24 \sqrt{3} \delta \psi \ .
\end{equation}

\begin{figure}
    \centering
    \includegraphics[width=\linewidth]{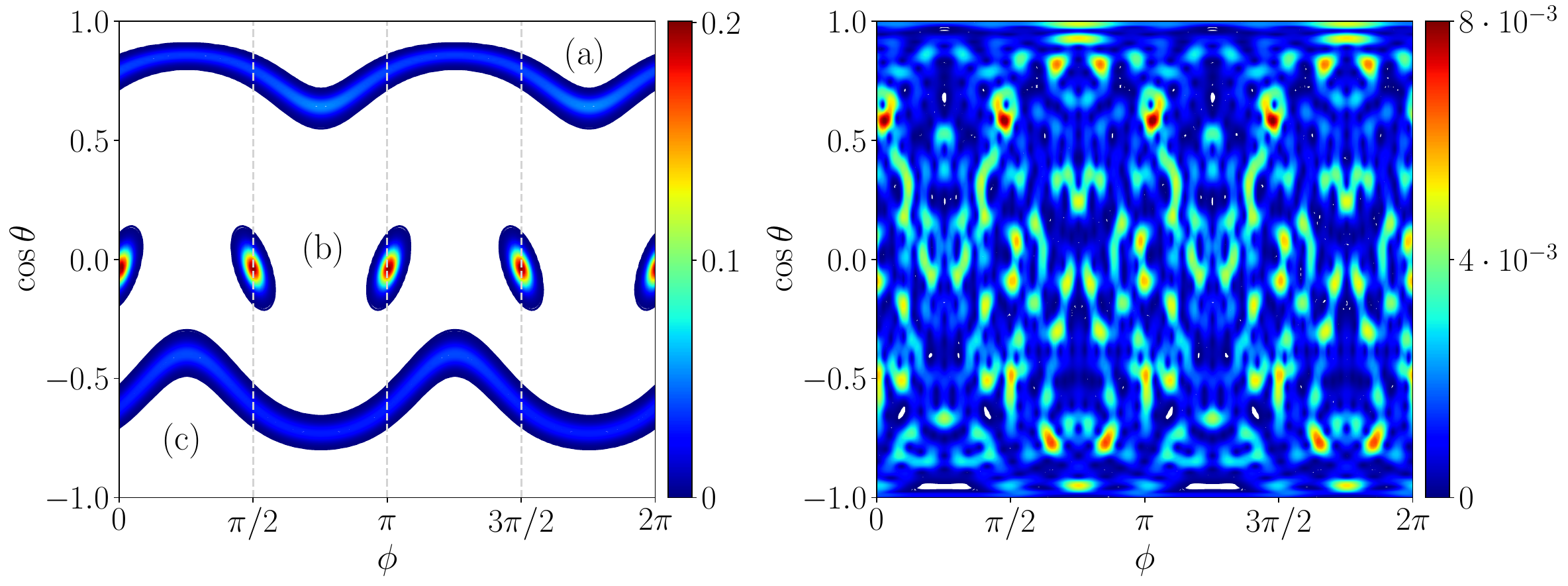}
    \caption{Color plot of the overlap $| \braket{\Omega_{\theta, \phi} | \eta_m}|^2$ between the spin coherent state $\ket{\Omega_{\theta,\phi}}$ and different Floquet eigenstates $\ket{\eta_m}$, corresponding to different phases, for $N=800$, $\psi = \pi/2 + 0.01$, $T=1$ (left) and $T= 10$ (right). While in the chaotic phase (right panel) the eigenstate has no structure, the eigenstate $(b)$ (left panel), which correspond to the DFTC phase with $p=4$, clearly exhibits the structure of a Bloch wave-function localized around the $\mathbb{Z}_4$ symmetric wells. The eigenstate $(b)$ (left panel) has maximum overlap with the spin coherent state corresponding to the initial conditions $\cos\theta = 0$, $\phi = \pi/2$.  Initial conditions localized around the eigenstates $(a)$ and $(c)$ (left panel) instead correspond to a quasi-periodic phase.
    }
    \label{fig:finiteN}
\end{figure}

\section{Finite size effects}
\noindent 
We now examine more closely the numerical results for finite $N$ (and $\alpha = 0$) presented in the main text. Let us notice that the fact that the modulus of the total spin $\mathbf{S}$ of the system is conserved allows us to restrict ourselves to the subspace with $\mathbf{S}^2 = s (s+1)$, with $s=N/2$, and thus to perform exact diagonalization up to large sizes ($N= 800$) \cite{RibeiroPRE2008,RussomannoPRB2017}. To visualize the eigenstates in this subspace, we introduce the spin coherent states\,\cite{auerbach2012interacting} 
\begin{equation}
    \ket{\Omega (\theta, \phi)}  = e^{-i \mathbf{n} \cdot \mathbf{S}} \ket{\uparrow},  
\end{equation}
where $\ket{\uparrow}$ is the eigenstate corresponding to the maximum projection of the spin along the $z$ direction and $\mathbf{n} = (\sin \theta \cos \phi, \sin \theta \sin \phi, \cos \theta)$. As
\begin{equation}
    |\braket{\Omega (\theta, \phi)|\Omega (\theta + \Delta \theta, \phi+ \Delta \phi)}| = \left( \sin \frac{\Delta \theta}{2} e^{-i \Delta \phi} \right)^{2s} \ ,
\end{equation}
while the $\lbrace \ket{\Omega (\theta, \phi)} \rbrace$  become orthogonal in the $N \rightarrow \infty$, for any finite $N$ they form an overcomplete basis for the Hilbert space. For various Floquet eigenstates $\ket{\eta_m}$, we estimated the projection $|\braket{\Omega (\theta, \phi)|\eta_m}|^2$. These eigenstates exhibit a quite different structure between the three different phases of the system: while no recognizable pattern is present in the chaotic phase (Fig.\,\ref{fig:finiteN}, right panel), in the quasi-periodic phase the eigenstate is localized in a connected region of the $(\theta,\phi)$ space (Fig.\,\ref{fig:finiteN},left panel, curves (a) and (c)), while in the $p=4$ DFTC phase it appears localized around four, $\mathbb{Z}_4$ symmetric, points (Fig.\,\ref{fig:finiteN},left panel: curve (b)). Let us notice that, given the initial condition chosen in the paper, in the $N \rightarrow \infty$ limit, the only eigenstate which contributes to the dynamics, will be the one with a non-zero overlap with the point $\theta = 0$, $\phi=0$, which in turn can correspond to each of the three phases. 

We can explain this behavior by noticing that in the classical limit, at stroboscopic times, the dynamics close to the resonance is dominated by the hopping between adjacent wells, each one localized around $\psi = k \psi_r$ (with $k=1, \cdots p-1$). Thus, for finite $N$, as the quantum effects are present, we expect the corresponding $\ket{\eta_m}$ to have the form of a Bloch superposition
\begin{equation}
    \braket{\Omega_{\theta, \phi} | \eta_m} = \sum^{p-1}_{k=0} e^{ 2 i \pi k/p} \ W_m (I,\phi- k \psi_r) 
\end{equation}
of the $p$ wavefunctions $W (I, \phi-k \phi_r)$. Those are connected by the Floquet propagator: 
\begin{equation}
    U_F W_m (I,\phi- k \psi_r) = e^{i \beta_m} W_m (I,\phi- (k+1) \psi_r). 
\end{equation}
As a consequence, we expect the modulus of this quantity to be localized around $\phi = k \psi_r$, i.e. the fixed points of the $p$-iterated evolution in the time-crystalline phase.

\section{Phase diagram beyond the fully-connected limit}\label{APP_sw}
In this appendix, we discuss the robustness of the Discrete Floquet Time Crystal (DFTC) phases, described in the main text, with respect to the inclusion of a perturbation on top of the fully-connected Hamiltonian
\begin{equation}\label{Hamiltonian_0}
    \hat{H}_0 = -\frac{J}{2 N} \sum_{i > j} \hat{\sigma}^i_x \hat{\sigma}^{j}_x + h(t) \sum_i \hat{\sigma}^i_z \ ,
\end{equation}
either obtained by replacing the all-to-all interaction with long-range, power-law decaying couplings, $J_{ij} \sim |i-j|^{-\alpha}$ (for $\alpha\neq 0$), or by the  inclusion of an extra short-range interaction term $\hat{U}= -\lambda\sum_{i} \sigma^x_i\sigma^x_{i+1} + \lambda(\sum_i \sigma^x_i)^2/N$ .
We treat the perturbation in the framework of non-equilibrium spin-wave theory (NEQSWT), originally developed in Ref.\,\cite{Lerose2018chaotic} and briefly sketched in the following (the interested reader may consult Refs.\,\cite{Lerose2018chaotic,Lerose2019impact} for further details). In the following, we will assume for simplicity that the systems is on a one-dimensional lattice. However, our calculation can be straightforwardly generalized to higher dimensions \,\cite{Lerose2019impact}.

\subsection{Review of non-equilibrium spin-wave theory}

The NEQSWT is useful to describe the unitary dynamics of systems whose Hamiltonian can be split in a fully-connected term and a perturbation, as
\begin{equation}\label{Hamiltonian}
    \hat{H} = - \frac{J}{4 N} (\tilde{\sigma}_x^{0})^2 + h(t) \tilde{\sigma}_z^{0} - \frac 1 N \sum_{k \neq 0} \tilde{\lambda}_k \, \tilde{\sigma}_x^k \tilde{\sigma}_x^{-k},
\end{equation}
where the Fourier modes are defined as $\tilde{\sigma}_\alpha^k = \sum_{j=1}^N e^{-ikj} \hat{\sigma}^j_\alpha$ and $k = 2\pi n/N$ for $n =0, \dots, N-1$, $N$ being the system size. The fully-connected term $\hat{H}_0$, including the first two terms of eq.\eqref{Hamiltonian} and equivalent to eq.\eqref{Hamiltonian_0}, generates the the dynamics of the magnetization $\vec{m} = \braket{\vec{\sigma}_0}$ described in the main text. In the limit of small couplings ${\tilde{\lambda}_k}$, we assume that the the dynamical excitation of "spin-waves" degrees of freedom, induced by the extra term in eq.\eqref{Hamiltonian}, is sufficiently small and can be treated perturbatively.

The perturbation theory is essentially implemented in three steps:
\begin{enumerate}
\item First, we express the dynamical evolution in a time-dependent, rotating reference frame $\mathcal{R}=\big(\mathbf{X}(t),\mathbf{Y}(t),\mathbf{Z}(t) \big)$, implemented via the unitary rotation $V\big(\theta(t),\phi(t)\big)=\exp\left(-i\phi(t) \sum_i \sigma_i^z/2\right)\exp\left(-i\theta(t) \sum_i \sigma_i^y/2\right)$, where the spherical angles $\theta(t)$ and $\phi(t)$ are fixed in such a way that the magnetization $\vec{m}(t)$ is aligned with the Z-axis in the new frame,
for any  $t>0$.
The spin operators transform accordingly:
\begin{equation}
V \hat{\sigma}^j_x V^{\dagger} = \mathbf{X} \cdot \vec{\sigma}_j \equiv \hat{\sigma}^j_X, \qquad 
V \hat{\sigma}^j_y V^{\dagger} = \mathbf{Y} \cdot \vec{\sigma}_j \equiv \hat{\sigma}^j_Y, \qquad 
V \hat{\sigma}^j_z V^{\dagger} = \mathbf{Z} \cdot \vec{\sigma}_j \equiv \hat{\sigma}^j_Z,
\end{equation}
and their evolution is described by the Heisenberg equation
\begin{equation}
    \frac{d}{dt}\hat{\sigma}^j_\alpha = - i [\hat{\sigma}^j_\alpha, \Tilde{H}]
\end{equation}
where the modified Hamiltonian $\Tilde{H} = H + i V \Dot{V}^\dagger$, includes a non-intertial, additional term $ i V \Dot{V}^\dagger = -s \vec{\omega}(t) \cdot \sum_j \vec{\sigma}_j$, where $\vec{\omega}(t) = (-\sin{\theta}\Dot{\phi}, -\Dot{\theta}, \cos{\theta}\Dot{\phi})$.

\item Assuming that the fluctuations, induced by the spin-waves and transverse to $\mathbf{Z}(t)$, are small, we expand the spin variables in the new frame $\mathcal{R}$ through the Holstein–Primakoff (HP) transformation\,\cite{auerbach2012interacting}:
\begin{equation}
\label{eq:HP_transf}
\tilde{\sigma}^i_X \simeq \sqrt 2 \hat{q}_i ,\quad     \tilde{\sigma}^i_Y  \simeq \sqrt 2 \hat{p}_i, \quad \tilde{\sigma}^i_Z  = 1 - (\hat{q}_i^2+\hat{p}_i^2-1) ,
\end{equation}
and keep only terms in the Hamiltonian \eqref{Hamiltonian} which are quadratic in the spin-waves modes $\Tilde{q}_k = \sum_r e^{-ikr}\hat{q}_r/\sqrt{N}$ and $\Tilde{p}_k =\sum_r e^{-ikr}\hat{p}_r/\sqrt{N}$.\footnote{Notice that in general the HP transformation is expressed in terms of ladder operators $b_i$,$b_i^\dagger$. Here we have $\hat{q}_i=(b_i+b_i^\dagger)/\sqrt{2}$ and $\hat{p}_i=(b_i-b_i^\dagger)/\sqrt{2i}$.}

\item The equations of motion for the time-dependent angles, $\theta(t)$ and $\phi(t)$, are obtained imposing self-consistently that
$\braket{\tilde{\sigma}_{0}^X(t)} = \braket{\tilde{\sigma}_{0}^Y(t)}=0$, leading to
\begin{equation}\label{collectivespindelta}
    \left\{
    \begin{split}
    \Dot{\phi} = & 4 J (\cos\phi)^2\cos\theta \big\{
    1 -\epsilon(t) \big\} + h(t) - 4 \delta^{qq}(t) \cos\theta\cos^2\phi+ 4 \delta^{qp}(t) \sin{\phi}\cos{\phi}
    \\
    \Dot{\theta} = & 4 J \sin{\theta}\cos{\phi}\sin{\phi}\{
    1 -\epsilon(t)\} -4\delta^{pp}(t) \sin{\theta}\sin{\phi}\cos{\phi} + 4 \delta^{qp}(t)\sin{\theta}\cos{\theta}\cos^2 \phi
    \end{split}
    \right.
\end{equation}
where $\delta^{\alpha\beta}(t) \equiv  \sum_{k\ne0} \tilde{\lambda}_k \Delta^{\alpha\beta}_k(t)/ (Ns)$ with $\alpha, \beta \in \{p,q\}$, is the "quantum feedback" by which the classical spin gets coupled to the corresponding spin-wave correlation functions, defined by
\begin{equation}
\Delta_k^{qq}(t) = \langle \Tilde{q}_{-k}(t)\Tilde{q}_k(t) \rangle ,
\qquad
\Delta_k^{qp}(t) = \langle \frac{\Tilde{q}_{-k}(t)\Tilde{p}_k(t) + \Tilde{p}_{-k}(t)\Tilde{q}_k(t)}{2} \rangle , \qquad
\Delta_k^{pp}(t) =\langle \Tilde{p}_{-k}(t)\Tilde{p}_k(t) \rangle
\end{equation}
and 
$\epsilon(t) = \sum_{k\neq 0} (\Delta_k^{qq}(t)+\Delta_k^{pp}(t)-1)/N$ is the spin-wave density of the dynamical excitations.
The equations of motion for the spin-wave correlations are straightforwardly derived from the Heisenberg equations for $\tilde{q}_k$ and $\tilde{p}_k$ and read  as
\begin{equation} \label{spinwavesequations_app}
\left\{
\begin{split}
 \frac{d}{dt}\Delta_k^{qq}= & 8 \tilde{\lambda}_k\cos{\theta}\sin{\phi}\cos{\phi}\Delta_k^{qq} + 8( J \cos^2 \phi -4\tilde{\lambda}_k \sin^2\phi)\Delta_k^{qp} \\
 \frac{d}{dt}\Delta_k^{qp}= & -4(J\cos^2 \phi - \tilde{\lambda}_k \cos^2 \phi \cos^2\theta)\Delta_k^{qq} + 4(J\cos^2 \phi -\tilde{\lambda}_k \sin^2\phi) \Delta_k^{pp} \\
\frac{d}{dt}\Delta_k^{pp}= & -8(J\cos^2 \phi - \tilde{\lambda}_k \cos^2 \phi \cos^2\theta) \Delta_k^{qp} -8\tilde{\lambda}_k\cos{\theta}\sin{\phi}\cos{\phi}\Delta_k^{pp}
\end{split}
\right.
\end{equation}
\end{enumerate}
We end up with the evolution equation \eqref{collectivespindelta}, describing the evolution of $\vec{m}(t)$, coupled to eq. \eqref{spinwavesequations_app}, describing the fluctuation induced by the spin-waves on top of the magnetization. Our approximation is valid as long the spin wave density is small, $\epsilon(t)\ll 1$.
In the fully-connected limit $J \to 0$ the evolution of the collective spin decouples from the fluctuations, and the equations \eqref{collectivespindelta} are equivalent to eq.\eqref{HamM}. In this case, the spin-wave correlators still have nontrivial dynamics, but the spin-wave density is conserved and always vanishes.

\subsection{Modified phase diagram}

Within the NEQSWT described above, we can investigate the dynamical phases detected by the order parameter
\begin{equation} \label{zetareg_2}
    \zeta^2 = \frac{1}{n_{\rm max}} \sum^{n_{\rm max}}_{n=0} \left( m_{x,n} (\psi + \delta \psi) -  m_{x,n} (\psi) \right)^2 \ ,
\end{equation}
beyond the fully-connected Hamiltonian limit.
In particular, we study the effect on the phase diagram (Fig.1 of the main text) due to the inclusion of a short-ranged perturbation to eq.\eqref{Hamiltonian_0} or to the substitution of the all-to all coupling with a power-law decaying term $J_{ij} \sim |i-j|^{-\alpha}$ (for $\alpha\neq 0$). The first case corresponds to $\tilde{\lambda}_k = \lambda \cos k$; in the case of long-range interaction with $0<\alpha<1$ and in the thermodynamic limit, we get a discrete spectrum with couplings \cite{Defenu2021}
\begin{equation} \label{LR_discrete_spectrum}
    \tilde{\lambda}_n = \lim_{N\to\infty} \tilde{\lambda}_{k= 2\pi n/N} = (1-\alpha)2^{1-\alpha} \int_0^{1/2} \frac{\cos (2\pi n s)}{s^\alpha}ds
\end{equation}

\begin{figure}
    \centering
    \includegraphics[width=1.1\textwidth]{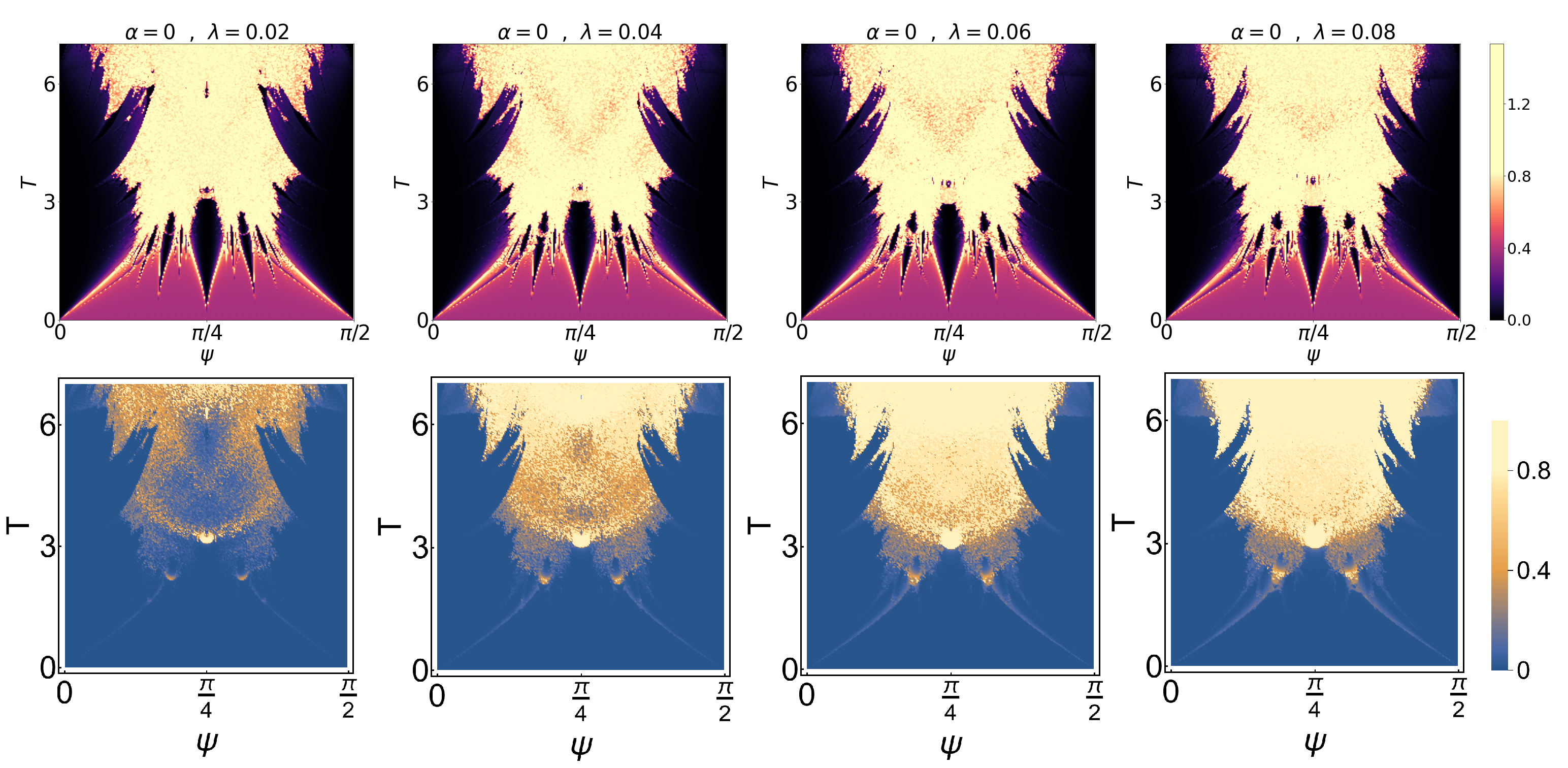}
    \caption{Phase diagrams resulting from the simultaneous integration of eq.\eqref{collectivespindelta} and \eqref{spinwavesequations_app}, for $\tilde{\lambda}_k = \lambda \cos k$. The momenta are discretized as $k = 2\pi n/N$, where $N=50$.
    \textbf{(Top)} Color plot of the order parameter $\zeta$ as a function of the amplitude $\psi$ and the period $T$ of the driving, 
    with $n_{max} = 300$, $\delta\psi=1.6\cdot 10^{-3}$.
    \textbf{(Bottom)} Color plot of the time-averaged spin-wave density $\overline{\epsilon}$, averaged up to a time $t=n_{max} T$, where again $n_{max} = 300$, for the same orbits computed in the phase diagrams on top.}    \label{fig:phase_diagram_NN}
\end{figure}

\begin{figure}
    \centering
    \includegraphics[width=1.\textwidth]{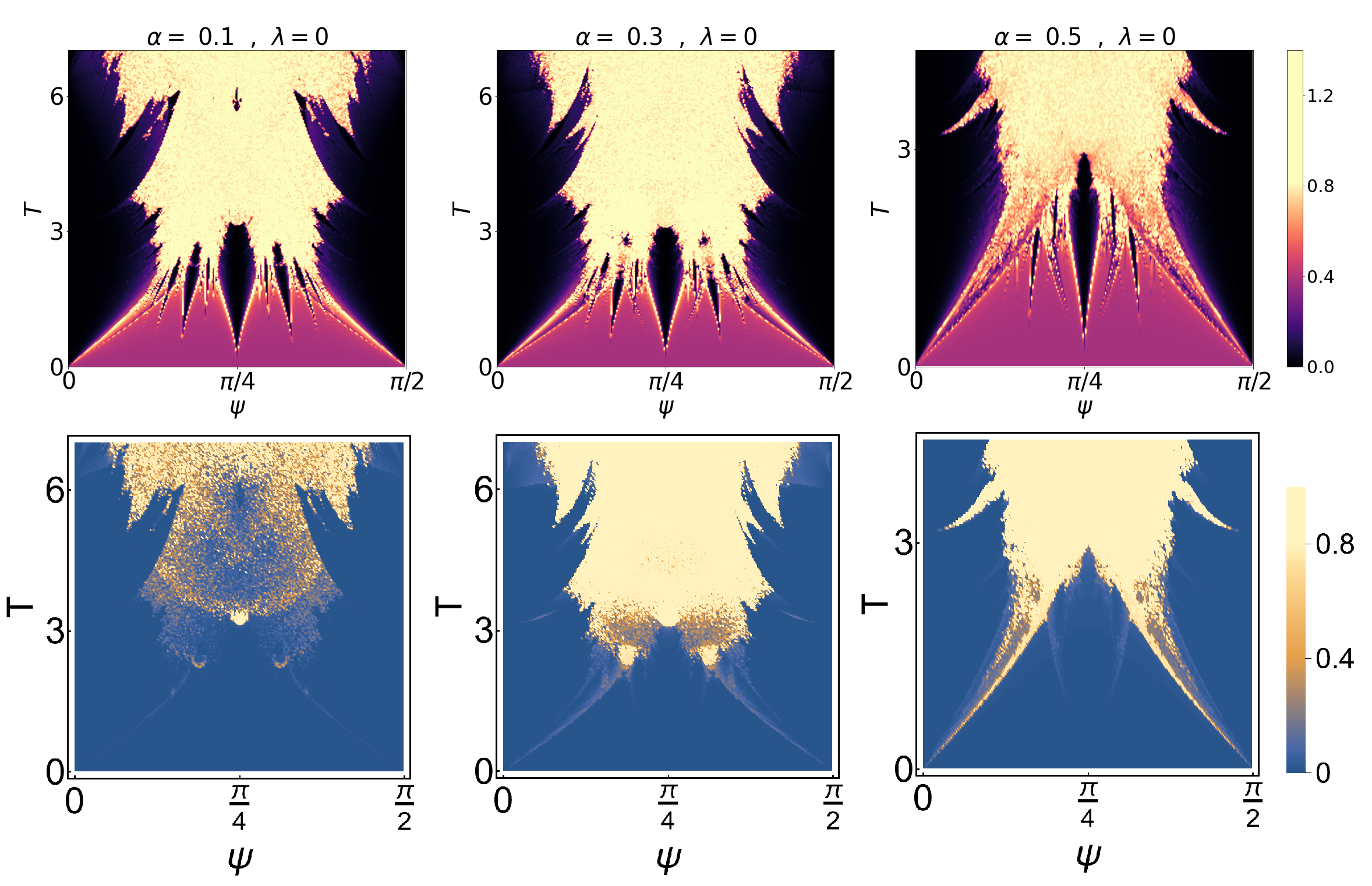}
    \caption{Phase diagrams resulting from the simultaneous integration of eq.\eqref{collectivespindelta} and \eqref{spinwavesequations_app}, for a discrete long-range spectrum $\tilde{\lambda}_n$ (see Sec.\ref{APP_sw}-B, for $n$ integer between $1$ and $50$.
    \textbf{(Top)} Color plot of the order parameter $\zeta$ as a function of the amplitude $\psi$ and the period $T$ of the driving, 
    with $n_{max} = 300$, $\delta\psi=1.6\cdot 10^{-3}$.
    \textbf{(Bottom)} Color plot of the time-averaged spin-wave density $\overline{\epsilon}$, averaged up to a time $t=n_{max} T$, where again $n_{max} = 300$, for the same orbits computed in the phase diagrams on top.}
    \label{fig:phase_diagram_LR}
\end{figure}

To compute $\zeta$, integrate simultaneously eqs.\eqref{collectivespindelta} and \eqref{spinwavesequations_app}, for fixed $\
\lambda$ and $\alpha$ and varying the period $T$ and the strength $\psi$ of the kick field $h(t)$, as done in the main text. Then we compute the longitudinal magnetization $m_{x,n} =\sin \theta(t_n) \cos \phi(t_n)$
,at stroboscopic times $t_n=nT$,
and straightforwardly obtain $\zeta$ from eq.\eqref{zetareg_2}.
In passing we notice that, in principle, when $\epsilon(t)\neq0$, the magnetization length is decreased by a factor $1-\epsilon(t)$. Here we are dealing with a "normalized" magnetization, whose length is $1$, in order to get a more transparent comparison with the results presented in the main text. This is not expected to qualitatively affect the phase diagram, as the order parameter $\zeta$ is sensitive only to the growth of the distance between neighbouring trajectories in parameter space.

The results, displayed in Figg.\,\ref{fig:phase_diagram_NN} and\,\ref{fig:phase_diagram_LR} (top) show that the DFTC are substantially stable beyond the fully-connected limit, in the range of parameters we considered: in particular, the we retrieve of DFTC islands at low $T$, while the isolated DFTC island around $\psi= \pi/4$, $T=6$ is shrinked by the perturbation and survives approximately up to $\lambda\simeq0.04$ and $\alpha\simeq0.3$.
In particular, the structure of the low-$T$ DFTC islands are not qualitatively altered by the perturbations, so that we can extend the analytic derivation done in Sec.\ref{APP_analytic estimation} to the dynamics studied in this section.
The stability of the phase-diagram with respect to spin-wave fluctuations can be understood also from Figg.\,\ref{fig:phase_diagram_NN} and\,\ref{fig:phase_diagram_LR} (bottom), where  the time-averaged spin-wave density $\overline{\epsilon} = \lim_{t\to \infty} \int_0^t\epsilon(\tau)d\tau/t$ is plotted for each trajectory: while the degree of excitation of the spin-waves stays small in the low-$T$ part of the phase diagram, where the quasi-periodic phase and the DFTC islands survive, $\overline{\epsilon}>1$ close to the chaotic orbits, where the system thus thermalizes\,\cite{Lerose2019impact} in presence of the spin-waves. This analyisis generalizes the one in Ref.\,\cite{PizziNatComm2021}, where the line at fixed $T=1$ was investigated.

\end{appendix}
\end{widetext}

\end{document}